\documentclass[sigconf]{acmart}

\copyrightyear{2020}
\acmYear{2020}
\acmConference[WWW '20 Companion]{Companion Proceedings of the Web Conference 2020}{April 20--24, 2020}{Taipei, Taiwan}
\acmBooktitle{Companion Proceedings of the Web Conference 2020 (WWW '20 Companion), April 20--24, 2020, Taipei, Taiwan}
\acmPrice{}
\acmDOI{10.1145/3366424.3384372}
\acmISBN{978-1-4503-7024-0/20/04}
\setcopyright{iw3c2w3}

\usepackage{microtype}
\usepackage{booktabs}                  

\begin{document}

\title{Toward An Interdisciplinary Methodology to Solve New (Old) Transportation Problems}

\author{Eduardo Graells-Garrido}
\orcid{0000-0003-0722-5969}
\affiliation{
    \institution{Barcelona Supercomputing Center (BSC)}
    \city{Barcelona}
    \country{Spain}
}
\affiliation{
    \institution{Universidad del Desarrollo}
    \city{Santiago}
    \country{Chile}
}
\email{eduardo.graells@bsc.es}

\author{Vanessa Pe\~na-Araya} 
\affiliation{
     \institution{Universit\'e Paris-Saclay, CNRS, Inria, LRI}
     \city{Paris}
     \country{France}%
}
\email{vanessa.pena-araya@inria.fr}

\begin{abstract}
The rising availability of digital traces provides a fertile ground for new solutions to both, new and old problems in cities.
Even though a massive data set analyzed with Data Science methods may provide a powerful solution to a problem, its adoption by relevant stakeholders is not guaranteed, due to adoption blockers such as lack of interpretability and transparency.
In this context, this paper proposes a preliminary methodology toward bridging two disciplines, Data Science and Transportation, to solve urban problems with methods that are suitable for adoption. The methodology is defined by four steps where people from both disciplines go from algorithm and model definition to the building of a potentially adoptable solution.
As case study, we describe how this methodology was applied to define a model to infer commuting trips with mode of transportation from mobile phone data.
\end{abstract}

%
%



\maketitle

\section{Introduction}
Cities are becoming more complex and traditional methods of data collection and analysis do not cope with such complexity.
The discipline of Transportation is particularly affected by these issues, not only due to city growth, but also to the arrival of new technologies and to changes in urban behavior. 
Another discipline, Data Science, has studied urban phenomena at previously unseen spatio-temporal granularity, mainly through the usage of mobile phone data~\cite{blondel2015survey}.
Arguably, even though both disciplines work on similar problems, they use a different domain language, resort to different data sources, and have different priorities.
Due to this gap between both fields, relevant institutions and transportation authorities/operators do not take advantage of the scalability, readiness, and granularity of Data Science-based models.
There is recognition on how a collaboration between them may deliver promising results~\cite{chen2016promises}, as such collaboration ``can be leveraged to evaluate urban planning policies that might affect economic growth, quality of life, environmental sustainability, and socioeconomic equity''~\cite{SanchezMartinez:2016wn}.
However, it has been noted that the application of data-driven technologies might not solve the complex problems of cities if they are not coupled with a range of other policies~\cite{Kitchin2014}, making the collaboration between data scientists and policy makers crucial. The question at hand is not the relevance of such collaboration, instead, given the gaps between disciplines, \textit{how} to conduct effective collaboration between them~\cite{Giest2017}.

The relevance of such collaborations becomes clear when considering the actual situation of cities. Consider Santiago (Chile), a city with almost 8 million inhabitants, with an integrated multi-modal transportation system. The city holds travel surveys every 10 years, a reasonable frequency rate in the previous decades, but inefficient in capturing the dynamics of the city today. Between the last survey held, in 2012, and 2020, the city has transformed in unexpected ways --- and not only in transportation terms. On the one hand, there are new ways of performing trips, such as ride-hailing applications, shared e-scooters, and ubiquitous routing services (e.g., Waze), which have changed how people make transportation choices. On the other hand, the population has acquired different habits. The last five years have seen an unprecedented international migration rate on the country, and migrants move differently through the city. Thus, the survey has its potential limited in planning and management of the transportation network, as the current transportation demand is unknown. Even though there is smart-card data that allows to count the number of trips in public transport, fare evasion is another problem that affects the transportation system, with current rates above 25\%.\footnote{\url{http://www.fiscalizacion.cl/}} This problem hides the actual public transport demand from planners. Authorities and operators are actively searching for a new type of solution that allows to complement the rich information of the travel survey with up-to-date observations of travel demand.

In this ever-changing context, frameworks such as Data Collaboratives~\cite{susha2017data} allow to define and kickstart interdisciplinary (Transportation, Data Science) and cross-sector (public, private) initiatives that could bring societal value. Once the project has started, how to create such value? How to ensure that the problem is solved in a way suitable for adoption?
Here we propose that this gap can be bridged through a four-step methodology that puts Data Science and Transportation to work around urban problems. Some of these problems are not new, but the complexities of modern cities makes the application of previous methods unfeasible; at the same time, new methods may lack needed qualities to be adopted in planning and policy making~\cite{castelvecchi2016can}.
The proposed methodology defines how both disciplines contribute to solve these problems through \emph{Algorithm Design}, \emph{Data-Driven Evaluation}, \emph{Interactive and Collaborative Evaluation}, and \emph{Knowledge Consolidation}.
%
In this paper we describe each step of the methodology, including the actors and concepts involved, as well as a case study of a method to infer the transportation mode share of the population in Santiago (Chile) using mobile phone data~\cite{graells2018inferring}.

By framing the development in the methodology, we obtained valuable insights for the building and adoption of Data Science solutions for urban problems.
We believe that such insights will support broader collaboration of Transport experts not only with Data Scientists but with citizens in general.

This paper is structured as follows. Section~\ref{sec:methodology} describes the four-step methodology. Section~\ref{sec:case_study} describes a case study for the problem of inferring the share of population that uses each mode of transportation in Santiago, Chile. Finally, Section~\ref{sec:conclusions} states our conclusions and next steps.

\begin{figure}[t]
    \centering
    \includegraphics[width=0.95\linewidth]{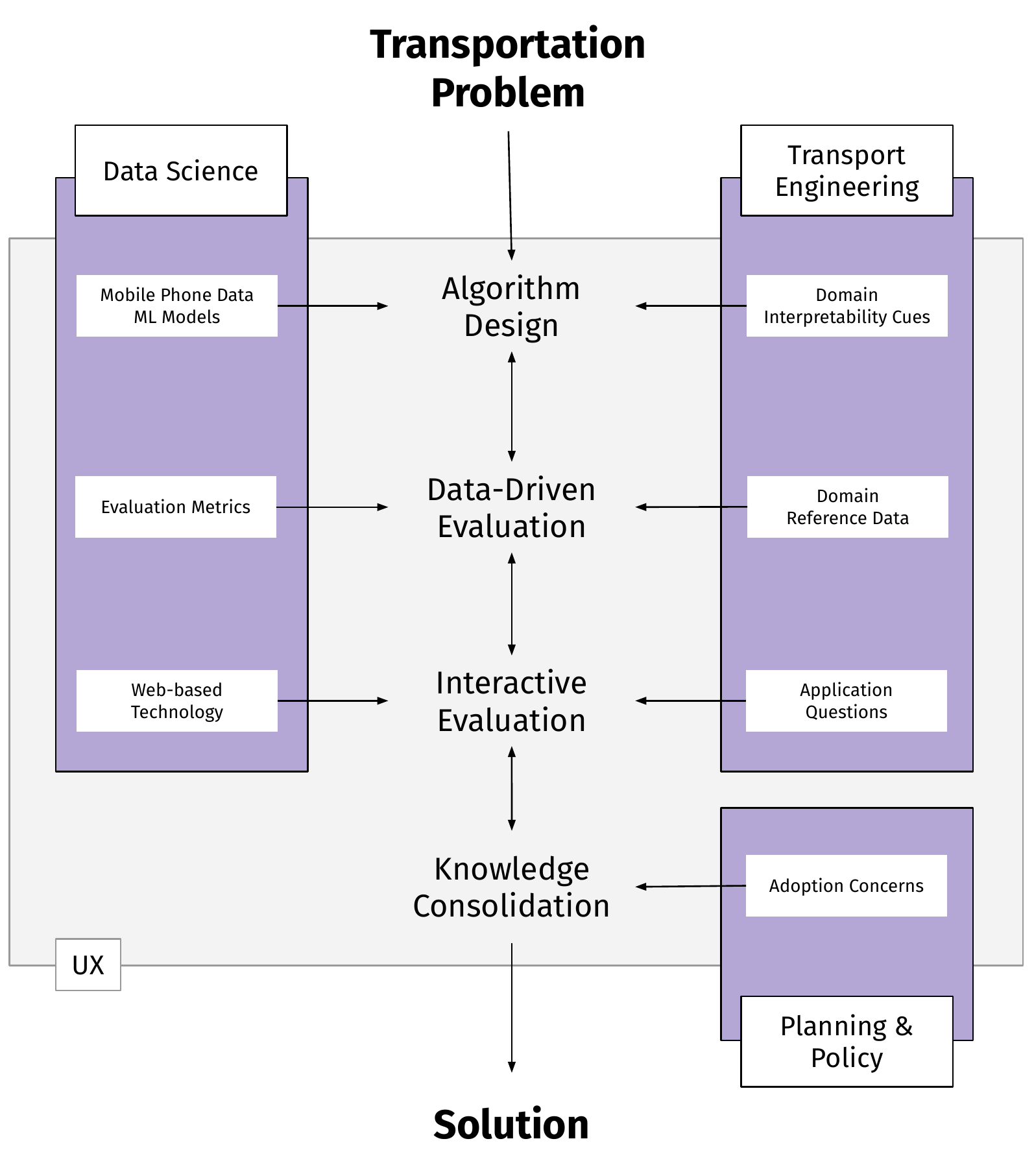}
    \caption{The four steps of the proposed methodology, with the relevant aspects contributed to each step from Data Science and Transportation. User Research (UX) is present in the whole process.}
    \label{fig:methodology}
\end{figure}

\section{Methodology}
\label{sec:methodology}

Our main goal is to encourage the adoption of non-traditional methods and data sets in new and old problems in transportation planning, where a ``new'' problem is likely to be a more complex and modern version of a known (``old'') problem. 
A \emph{transportation problem} is defined as a set of requirements to be solved by urban planners and/or policy makers, through the usage of data, that may be traditionally used in the problem domain, or may be non-traditional, such as mobile phone data. The latter type of data may require an advanced analysis to be used in the context of the problem, and thus, the need to use Data Science methods in designing and implementing a solution. This raises several challenges, because a research system with the purpose of having impact on public policy needs to provide the qualities of transparency and interpretability~\cite{krause2016interacting,castelvecchi2016can}.

This methodology is designed to be applied \emph{after} a transportation problem has been identified. As such, it assumes that there are several actors involved with the development and adoption of the potential solution. 
In this context, the methodology is comprised by the following steps: 1) \emph{Algorithm Design}, 2) \emph{Data-Driven Evaluation}, 3) \emph{Interactive and Collaborative Evaluation}, and 4) \emph{Knowledge Consolidation}. These four steps are sequential in appearance, however, the framework could be iterative, depending on the development strategy followed in each project. 
Each step receives specific input and tools from each discipline, and all steps use tools from user experience research methods to enhance communication between people from different backgrounds~\cite{kuniavsky2003observing}. This is important, for instance, to ensure that the technical language differences between disciplines converge into opportunities rather than limitations during the development of the solution.

A visual depiction of the methodology is shown on Figure~\ref{fig:methodology}. In the following, we describe each step.

\subsection{Algorithm Design}
The purpose of this step is to enable a joint implementation of a solution to the problem. 
Each discipline has their own processes to solve a problem through model building and algorithm design. 
Hence, on the one hand, Data Science contributes \emph{new data sources} that may contain proxy information to solve a problem, as well as \emph{machine learning models} to find patterns and make predictions. On the other hand, Transportation contributes \emph{interpretability cues}, i.e., aspects of the problem (and its potential solution) that cannot be hidden in a black-box method. For instance, when designing a model (or a model pipeline), data scientists may optimize performance according to the model interpretability by transportation practicioners in addition to accuracy. 

Then, to reach a consensus in what interpretability means for actors from both disciplines, pilot co-design workshops could be done to define a common definition, and find a common language between disciplines.

The output of this step is an algorithmic model with interpretable results that has the potential to solve the problem at hand.

\subsection{Data-Driven Evaluation}
A typical evaluation of the proposed model from the previous step would compare its performance with a reference data set with metrics such as accuracy, precision, and recall. 
However, this is hard for two reasons.
First, the availability of ground-truth data is not guaranteed. Transportation experts may have data sources that describe urban phenomena, such as travel surveys, however, such sources may be outdated, or may be sparser than the data set available for the project, such as digital traces from mobile phone data. 
Second, typical evaluation metrics do not necessarily measure the transparency of a model~\cite{krause2016interacting}. 
Questions addressed in the co-design process include:
\emph{How does the model behave with validation data?} \emph{What is the extent of the validation data, what are the intrinsic differences between the validation data and the non-traditional data?}
\emph{How each interpretable feature from the model explains the phenomena?}

In this step we propose to co-design the answers between data scientists and transportation practicioners. 
For instance, methods to explain how any model works already exist~\cite{NIPS2017_7062}, thus, the main challenge is to find the relationship between these explanations and transparency, as well as balancing the trade-off between transparency and accuracy metrics. 

The output of this step is a model that has been validated through Data Science and Transportation criteria, and that has qualities of interpretability and transparency.

\subsection{Interactive and Collaborative Evaluation}
At this point in the methodology, a potential solution exists, as the output of the previous step. 
However, it still needs to answer the following question: \emph{Would domain experts use the solution in an applied context?}
To find the answer, we propose to use visualization in collaborative environments for such validation, particularly with large displays.
Collaborative environments can facilitate interdisciplinary communication and development of shared ideas.
In the context of data analysis, information visualization and big displays are powerful tools to enable such environments. Together they put together a space where domain experts engage with Data Science experts, providing feedback, asking new questions, framing results; enabling a continuous process of analysis~\cite{Ball:2007:MIP:1240624.1240656, Endert:2011:VES:1992917.1992935}.
In our context, visualization is a tool valued by transportation researchers~\cite{cheng2013exploratory}, and transportation has been a recurring topic in visual analytics~\cite{andrienko2013visual}.
Taking advantage of such an environment requires the design and implementation of technology, for instance, such as Web-based visualization frameworks, and the design of user research instruments to faithfully capture the interactions between experts. This combination of technology and user research allows to derive an aswer to whether the proposed method in the project, although already validated in analytical terms, would be actually used in applied contexts. 

The output of this step is a model that has been validated in theoretical and applied contexts.

\subsection{Knowledge Consolidation}
In this stage, decision makers need to assess what is needed for the proposed solution to be adopted, by identifying concerns relevant for the adoption of the project. It is a complex scenario, as decision makers have to look into at many more dimensions of the problem that the technical actors involved are able to see.
Potential concerns include: definition of intellectual ownership and exploitation agreements with the data provider; the cost of accessing, storing, and updating the data; the compliance with privacy regulations; and the limits of population representativeness in the available digital traces, among others. Some of these concerns can be addressed by iterating over the previous steps, while others require actions outside of the scope of this methodology, such as political agreements between private and public actors~\cite{susha2017data}. 

In this step the output is a communication device built considering insights and technical aspects of the solution, with information gathered through user research and communication practices. The audience of this device are the aforementioned decision makers. Its content is the consolidated knowledge of the project, aiming to be a useful tool to push for adoption by decision makers.

\null

In summary, these steps provide guidelines on how to iteratively design and validate an interdisciplinary solution to a transportation planning problem. In the next section we describe how this methodology was applied in an existing project.

\begin{figure}[t]
    \centering
    \includegraphics[width=\linewidth]{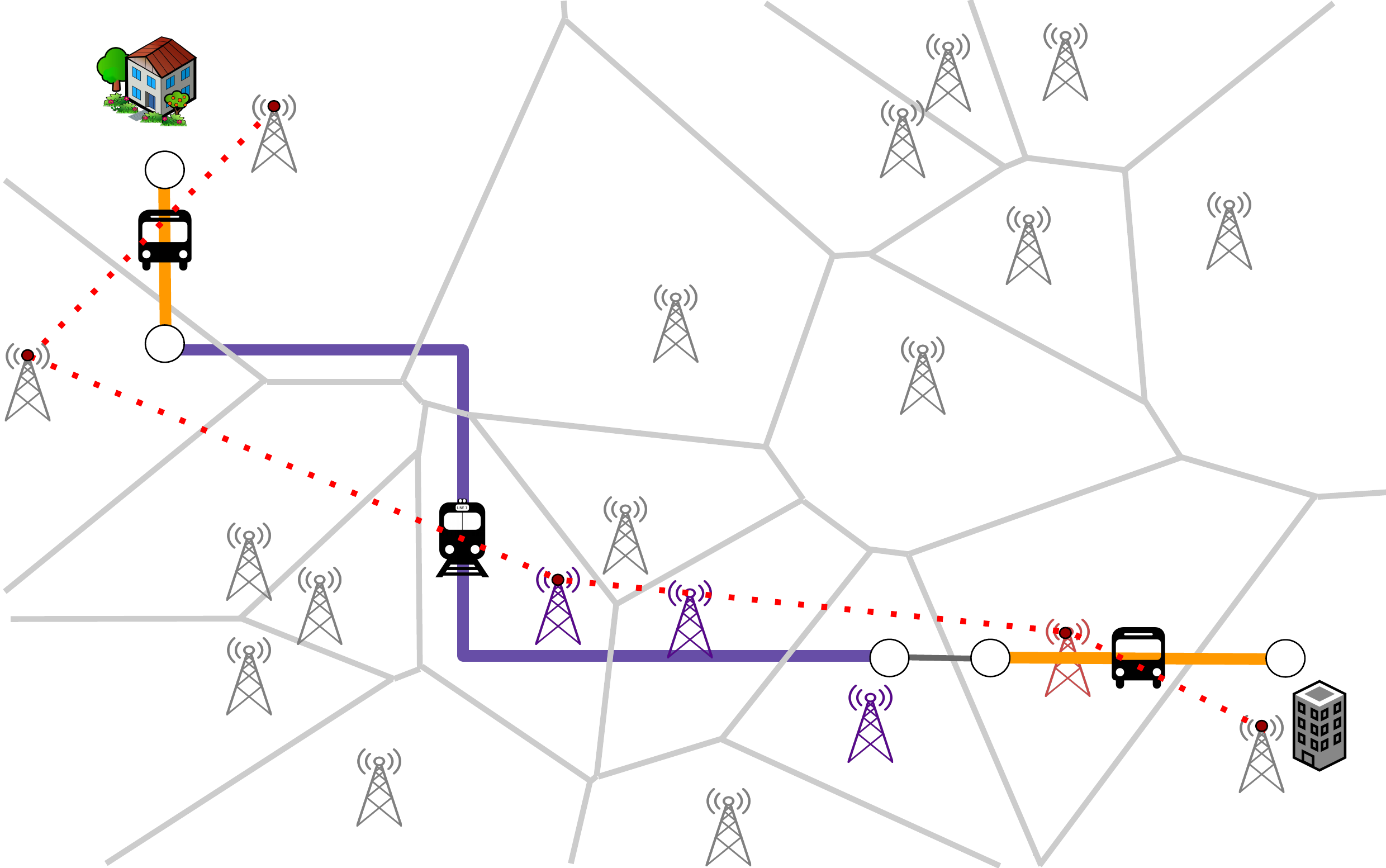}
    \caption{Representation of a commuting trip in mobile phone data. The sequence of towers connected to during a trip may only resemble the real trajectory.}
    \label{fig:xdr}
\end{figure}

\begin{figure*}[t]
    \centering
    \includegraphics[width=\linewidth]{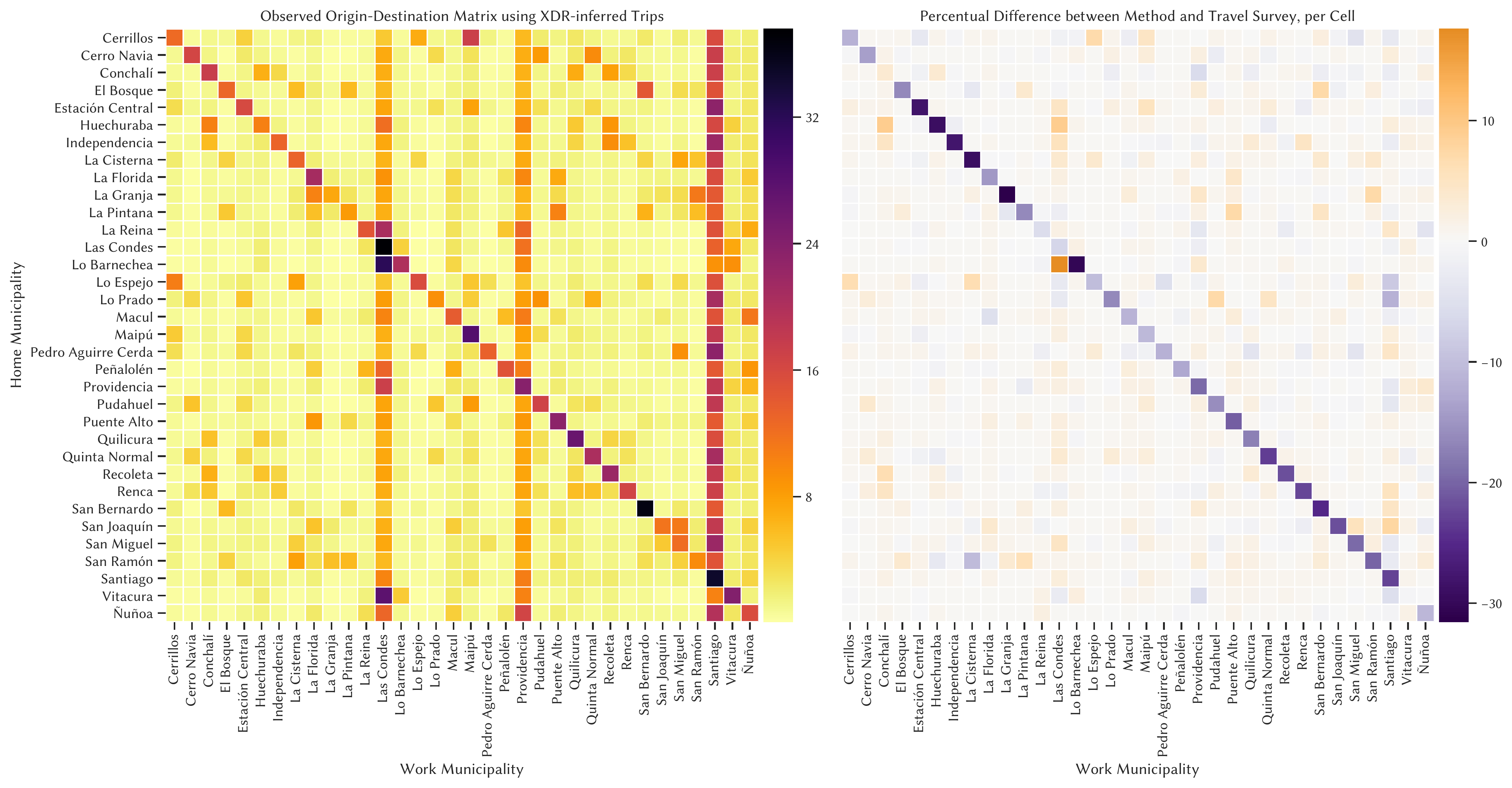}
    \caption{A co-designed visualization. The left matrix is row-normalized and depicts the number of trips from/to all areas of the city, as inferred from mobile phone data. The right matrix shows the difference between the left matrix and an analogously built matrix from a travel survey. A divergent color palette allows to identify critical cells that can be analyzed manually.}
    \label{fig:od_matrices}
\end{figure*}

\section{Case Study}
\label{sec:case_study}

In this section we describe how the methodology was applied to the problem of inferring the mode of transportation share in Santiago (Chile) from anonymized mobile phone data~\cite{graells2016day,graells2018inferring}. This problem is traditionally solved through the collection of travel surveys, which provide rich information used to design changes and additions into the transportation network. At the same time, surveys are expensive and require efforts of collecting and analyzing data. Conversely, mobile phone data is already generated and stored by telecommunication operators, and they contain proxy information of how people move through the city (see Figure~\ref{fig:xdr}). This situation prompted a collaborative effort between Telef\'onica Chile, the largest telco. in the country, and the Data Science Institute, who provided data know-how and research, and transportation practicioners from the Office of the Secretary of Transportation in Chile. We followed the methodology described in Section~\ref{sec:methodology}. Note that the methodology was drafted initially, however, its current design benefited from the implementation of this case study.

First, in the \emph{Algorithm Design} step, we jointly defined that the problem could be described as a soft clustering problem, where each mode of transportation is a cluster. After testing the accuracy of several clustering methods, we found that a model known as Topic-Supervised Non-Negative Matrix Factorization~\cite{macmillan2017topic} was commonly assessed as an interpretable  model due to is simple mathematical definition. It is a semi-supervised model that allowed us to incorporate knowledge of the transportation network into the learning process. 
We organized two pilot workshops, one at a university and one at the premises of the mobile phone company. There we converged the model description to domain-language from both disciplines, and defined how to incorporate domain knowl\-edge into the model. We learned that the model is not only interpretable because of its mathematical definition, but also because the methods employed in finding a solution are known in Transport Engineering, which commonly solves optimization problems, such as optimal bus fleet deployment.

\begin{figure}[t]
    \centering
    \includegraphics[width=\linewidth]{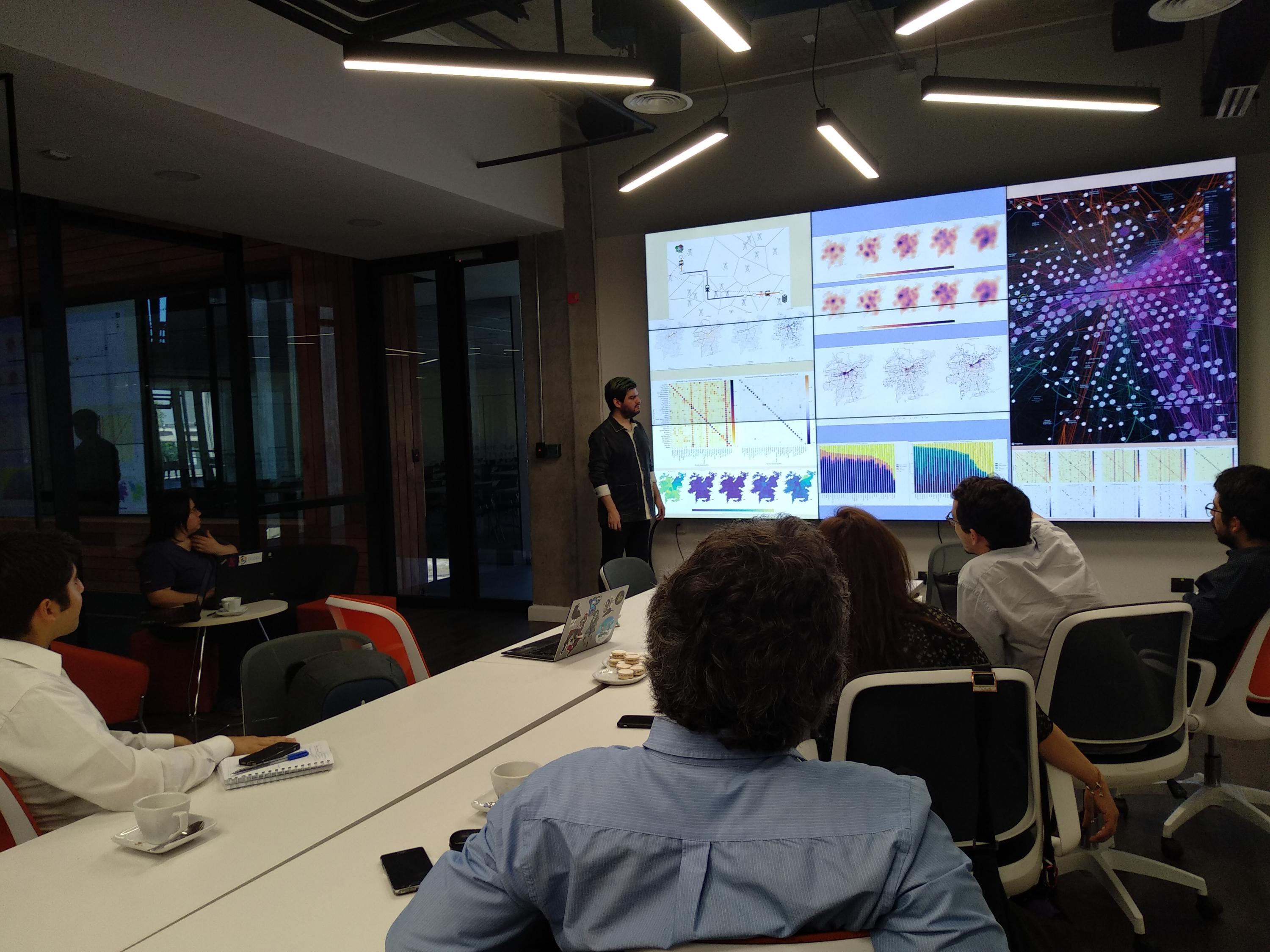}
    \caption{Interactive evaluation of the proposed model with domain experts in a collaborative environment.}
    \label{fig:session_pic}
\end{figure}

\begin{figure*}[t]
    \centering
    \includegraphics[width=\linewidth]{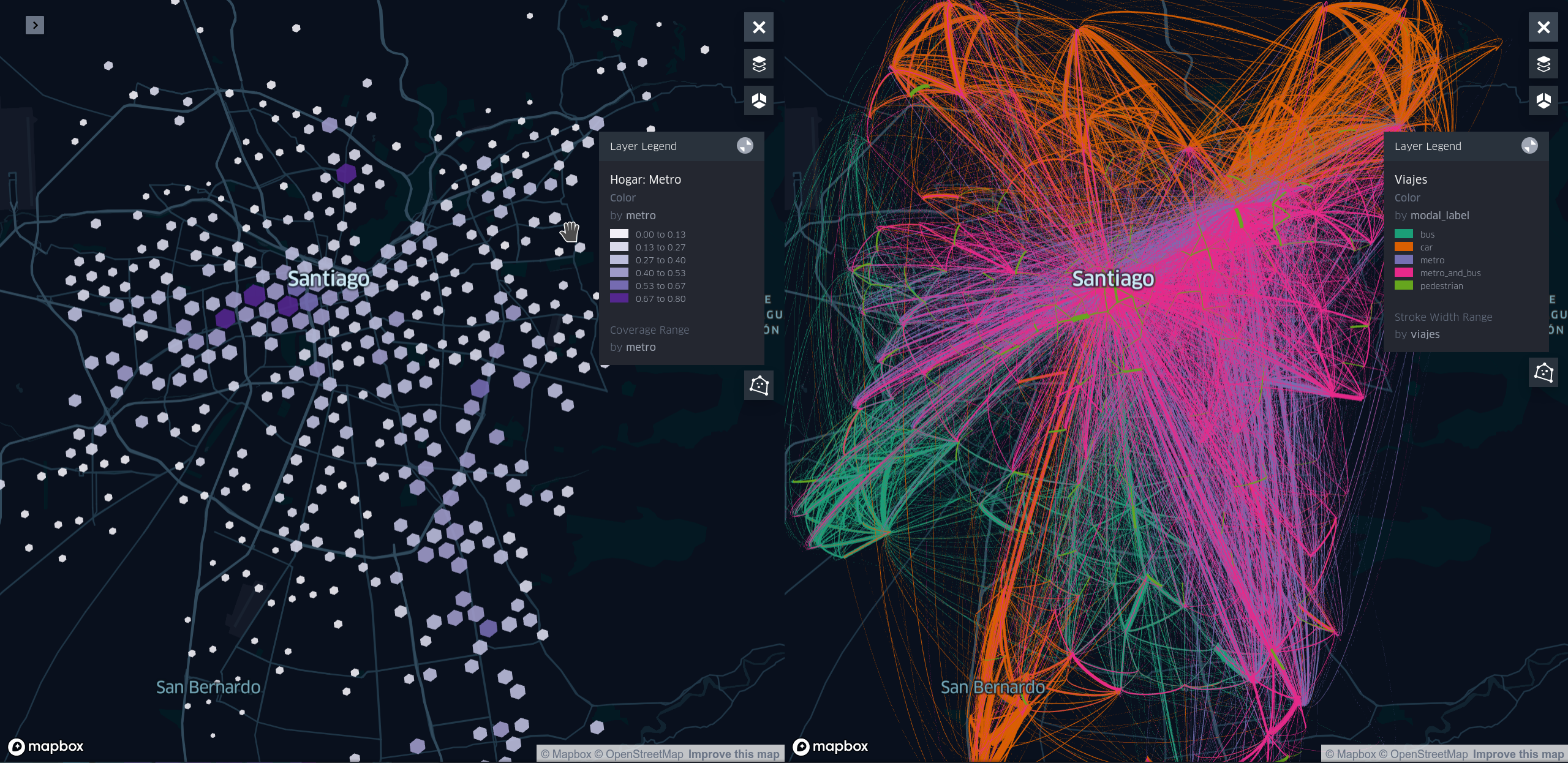}
    \caption{Interactive visualization used by the domain experts in the interactive session. This view shows a coordinated view with the spatial distribution of metro users (left), and the entirety of flows estimated by the model, colored according to primary transport mode per flow.}
    \label{fig:kepler}
\end{figure*}

In the next step, \emph{Data-Driven Evaluation}, we compared our work with a travel survey, as suggested by urban planners. Even though the survey was from 2012, and that our data was from 2016, domain experts mentioned that the comparison was still relevant, and provided insight of which aspects they would evaluate in the model. Since they work with aggregated flows in the city, they were more interested in the aggregated results of the model. Thus, they needed to assess how similar an Origin-Destination (OD) matrix from mobile phone data was to the corresponding one from a travel survey. We agreed that a correlation coefficient would be enough for them to assess the quality of the model, having into account that some OD pairs in the travel survey do not exist and should not be considered in the correlation.

One of our goals when we started the pilot workshops was to understand the importance given to interpretability and transparency in machine learning models. We expected that visualization could aid in this aspect. However, there was an additional factor that was more important for domain experts: the meaning of \emph{error}. We learned that a model with considerable error may be acceptable for experts, provided that this error is understood, and that measures to correct it can be taken.
For instance, if the model incorrectly predicts mode of transportation for specific areas of the city, small, targeted surveys could be conducted, and such results would be apt to include in a new iteration of the model definition.

From this discussion we co-designed a visualization to compare OD matrices (see Figure~\ref{fig:od_matrices}). We applied this design to evaluate the OD matrices obtained for each mode of transportation (see reference~\cite{graells2018inferring} for its application to all modes of transportation).

In the third step, \emph{Interactive and Collaborative Evaluation}, we aimed at facilitating the collaborative analysis of the data-driven model and the communication among experts. To do so, we performed an interactive analysis session using large displays with Web-enabled technologies.
The large display was composed by a 3 $\times$ 3 array of Samsung UD55E-B video wall, using the SAGE2 platform.\footnote{\url{https://sage2.sagecommons.org}} 
The results from the previous stages were visualized in two ways. Firstly, using the co-designed visualizations in real time on the large display through a Jupyter lab extension for SAGE2.\footnote{\url{https://github.com/AndrewTBurks/jupyterlab\_sage2}}
Secondly, through an interactive visualization implemented in the geo-visualization framework \url{Kepler.gl}\footnote{\url{https://kepler.gl/}} which allowed an appealing and highly dynamic representation of the data. 
The room was arranged as a round table in which all participants could see each other and interact in a comfortable way. Participants were able to connect their mobile devices to the screen and use an interactive pointer within the system (see Figure~\ref{fig:session_pic} for the whole setup, and Figure~\ref{fig:kepler} for a detailed view of the interactive visualization), as a way to enable single and collaboratively built queries.
Four potential end-users attended the interactive session: two policy makers from the Ministry of Transportation and two practitioners, one from a research center and one from the subway operator. 

The most fruitful information was extracted from the conversation between experts during the session. From their perspective, this kind of model could be valuable, however, they questioned how the model outcomes could be integrated in their own pipeline of tools, in both aspects, operational and methodological. 
On the one hand, they need more estimations of the data and model bias and errors, which were still not provided by the visualization, nor the model. This may be due to the expectations that Data Science solutions create on the domain experts, for instance, they requested information about the first and last stages of trips. The model only identifies whether people use multi-modal transportation, not necessarily the order of modes in their commuting trips.
On the other hand, they declared the importance of including socio-demographic information about the people from whom this data was generated. However, we were not able to provide a temporary solution due to the anonymized nature of the data.
None of these insights were evident on the previous steps of the development, implying that, when thinking about applying the proposed solution, the criteria is different when designing a research solution. 

Finally, in the \emph{Knowledge Consolidation} step we reasoned that there are other additional attributes that a data science-based system needs for adoption. \emph{Interpretability} and \emph{transparency} are the two most common, however, we learned that \emph{interoperability} is equally or even more important in transportation.
Interoperability is seen by the domain experts as a way of updating and calibrating already existing models and systems, which the domain experts trust (and won't stop using in the short term), and generating fine-grained input data for their systems.
Indeed, transportation experts are looking for ways to modernize their tools and methodologies instead of replacing them. 
Hence, an adoption concern is not whether the model works or not (it does), is its feasibility of integration with existing tools used by planners. We transferred the implementation to the telecommunications operator as a software package, and we socialized the results with the transportation authorities through meetings. 

We have learned that, to encourage adoption in a big city as Santiago, the authorities are willing to do data-driven studies in smaller cities. We also identified future lines of research that could have a positive impact on the interaction between disciplines. For instance, we are currently working on new ways of visualizing multivariate flows in the city~\cite{perez2019visualizing}.

\section{Conclusions}
\label{sec:conclusions}

In this work we defined a methodology to solve new (old) problems in transportation, through the collaboration of domain experts and data scientists. This methodology was applied in the design and implementation of a model to work on a current problem, the inference of the mode of transportation share in a city of 8 million inhabitants. It allowed us to identify key contributions in the model that would enhance its interpretability and transparency, as well as to identify application and adoption concerns that need to be addressed before a concrete usage of it. 

One of the limitations of this paper is the preliminary aspect of the methodology. We have applied it to a specific problem, with positive results, but it still remains a challenge to see how it would behave in other projects that are not so tied to its definition. Moreover, it is not clear whether the problem could have been solved effectively using a different methodology. However, in contrast with common user studies, the number of available end-users for this type of task is rather limited, thus, a set of guidelines that help to build trust and to improve the qualities of the model is a valuable asset.
As future work, we aim to continue studying how machine learning, visualization, and user research, could impact the adoption of non-traditional data sources for urban planning and policy making.
Doing so would reduce knowledge gaps between disciplines. 
We expect to make explicit the participation of citizens to understand how privacy plays a role in these projects. 

\begin{acks}
We wish to thank the Atelier Team at Inria Chile (\url{http://inria.cl/}) for help in organizing and holding the interactive session, specially Celeste Bertin for her technical support. We also thank Francisca Ar\'evalo for help in organizing and designing the pilot workshops, and Karina Flores for interviewing experts. Finally, we thank Viviana Mu\~noz from the Secretary of Transportation for helping to build bridges between disciplines. E. Graells-Garrido was partially funded by CONICYT Fondecyt de Iniciaci\'on project \#11180913.
\end{acks}


\bibliographystyle{ACM-Reference-Format}
\bibliography{main}
\end{document}